# Improvement of local critical current density of $RE$Ba$_2$Cu$_3$O$_{7-\delta}$ by the increase in configurational entropy of mixing at the $RE$ site

Aichi Yamashita*, Yuta Shukunami, Yoshikazu Mizuguchi*

*Department of Physics, Tokyo Metropolitan University, Hachioji 192-0397, Japan*

*Corresponding authors: Aichi Yamashita (aichi@tmu.ac.jp), Yoshikazu Mizuguchi (mizugu@tmu.ac.jp)

**Abstract**

$RE$Ba$_2$Cu$_3$O$_{7-\delta}$ ($RE$123, $RE$: rare earth) is one of the high-temperature superconductors with a transition temperature ($T_c$) exceeding 90 K. Because of its high $T_c$ and large critical current density ($J_c$) under magnetic fields, $RE$123 superconductors have been expected to play a key role in superconductivity application. To accelerate application researches on $RE$123-based devices, further improvements of $J_c$ characteristics have been desired. In this study, we investigated the effects of high-entropy alloying at the $RE$ site on the superconducting properties, through the measurements of local (intra-grain) $J_c$ ($J_c^{\mathrm{local}}$) by a remanent magnetization method. We found that $J_c^{\mathrm{local}}$ shows a trend to be improved when four or five $RE$ elements are solved at the $RE$ site, which results in high configurational entropy of mixing ($\Delta S_{\mathrm{mix}}$). Because high-entropy alloying can improve $J_c^{\mathrm{local}}$ of $RE$123 superconductors by modification of the $RE$ site composition and $\Delta S_{\mathrm{mix}}$, and the technique would be applicable together with other techniques, such as introduction of nanoscale disorders, our entropy-engineering strategy introduced here would be useful for development of $RE$123 superconducting materials available under high magnetic fields.

**Introduction**

Since the discovery of high-transition temperature (high-$T_c$) Cu-oxide superconductor[1], various kinds of Cu-oxide superconductors have been discovered[2-5]. Among them, $RE$Ba$_2$Cu$_3$O$_{7-\delta}$ ($RE$123, $RE$: rare earth) in a thin-film form is a promising material for high-field superconductivity application because of its high $T_c$ exceeding 90 K and high critical current density ($J_c$) under magnetic fields[6]. To improve $J_c$ of $RE$123 films, nanoscale disorders, such as nanoparticles, nanocomposite structure, defects etc., were introduced[7-10]. However, the current record of $J_c$ of $RE$123 films are far from the limit expected for an ideal $RE$123 materials[11]. To achieve higher $J_c$ in $RE$123 materials, further development of the method for $J_c$ engineering is needed. Having considered the structural and physical properties of various Cu-oxide superconductors, we find that the $RE$123 system is relatively difficult to utilize in superconductivity applications because it contains structural and compositional fluctuations. High-$T_c$ and high-$J_c$ superconductivity of $RE$123 is generally observed in the orthorhombic structure with the space group of *Pmmm* (#47)[12]. Decrease in oxygen amount in $RE$123 in the blocking layer results in a decrease in hole carriers and suppression of high performance of $RE$123. Recently, improvement of $J_c$ in $RE$123 film was achieved by over-doping of holes[13]. The improvement of $J_c$ by chemical-composition tuning would be a desired progress because that can be applied together with the nanoscale fabrication techniques mentioned above. Here, we show another strategy to improve local (intra-grain) $J_c$ ($J_c^{local}$) by introducing high-entropy-alloy-type (HEA-type) $RE$ site in $RE$123 grains.

HEA is an alloy containing five or more elements with a concentration range between 5 to 35 at% and hence has a high configurational entropy of mixing ($\Delta S_{mix}$), which is defined as $\Delta S_{mix} = -R \Sigma_i c_i \ln c_i$, where $c_i$ and $R$ are compositional ratio and the gas constant, respectively[13,14]. Although the field of HEA had mostly focused on structural materials for the use under extreme conditions, various functionalities have been found in HEAs[14,15]. In 2014, superconductivity was observed in a HEA, Ti-Zr-Hf-Nb-Ta[16]. Although the expected pairing mechanisms of superconductivity for the HEA was conventional type, the unique structural and compositional character were welcomed in the field of new superconducting materials. As reviewed in Ref. 17 and 18, many HEA superconductors were discovered after the first discovery by Koželj et al. Since 2018, we have developed

HEA-type superconducting compounds, in which the HEA concept was applied to complicated compounds having two or more crystallographic sites[19]. Comparing the HEA effects for superconductors with various crystal structural dimensionality, we found that the disordering effects introduced by the HEA-type site in layered system ($BiS_2$-based superconductor)[20] and quasi-two-dimensional system (tetragonal $TrZr_2$, $Tr$: transition metals)[21,22] does not suppress its original $T_c$ in pure phases. In contrast, in three-dimensional systems (NaCl-type metal tellurides[23–25] and A15 niobium-based compounds[26]), $T_c$ of HEA-type phases was clearly lower than that for pure phases. Therefore, in a two-dimensional crystal structure, introduction of HEA-type site does not negatively work on $T_c$ of the superconductor. As a result, we previously reported the synthesis of HEA-type $RE$123 polycrystalline samples and reported that the increase in $\Delta S_{mix}$ does not suppress superconducting properties including $J_c$[26]. Let us remind that the technique of mixing $RE$ elements have been used for improvement of $J_c$ in some $RE$123 materials[27]. To the best of our knowledge, however, superconducting properties including $J_c$ for $RE$123 with remarkably high $\Delta S_{mix}$ at the $RE$ site was examined in Ref. 26 for the first time. In this study, we expanded the study and synthesized various samples of $RE$123 using lighter $RE$ elements including Dy, Ho, Yb, and Lu. Here, we show that tuning $\Delta S_{mix}$ at the $RE$ site could improve intra-grain $J_c$ ($J_c^{local}$) of $RE$123 superconductors.

**Materials and methods**

*Sample preparation and characterization*

Polycrystalline samples of $REBa_2Cu_3O_{7-\delta}$ ($RE$: Y, La, Nd, Sm, Eu, Dy, Ho, Yb, Lu) were prepared by solid-state reaction in air, as described in Ref. 26. Powders of $Y_2O_3$ (99.9%), $La(OH)_3$ (99.99%), $Nd_2O_3$ (99.9%), $Sm_2O_3$ (99.9%), $Eu_2O_3$ (99.9%), $Dy_2O_3$ (99.9%), $Ho_2O_3$ (99.9%), $Yb_2O_3$ (99.9%), $Lu_2O_3$ (99.9%), $BaCO_3$ (98%), and CuO (99.9%) were used for the synthesis. To obtain the best superconducting properties of each sample, two-step or three-step sintering was performed. The raw chemicals with a nominal compositional ratio of $RE$ : Ba : Cu = 1 : 2 : 3 were well mixed and pelletized with a diameter of 1 cm. The first sintering condition was 930°C for 20 h, followed by furnace cooling. For the second sintering process, the sample was ground, mixed, pelletized in the same manner as the first one, and heated at 930°C for 8 h and 350°C for 18 h, followed by furnace cooling. The third sintering was performed for samples which showed lower $T_c$ and shielding volume fraction after the second sintering. The condition of the third sintering was 930°C for 8 h, 350°C for 18 h, and 175°C for 12 h, followed by furnace cooling.

Powder X-ray diffraction (XRD) patterns were collected on MiniFlex-600 (RIGAKU), equipped with a D/tex-Ultra high-resolution detector, with a Cu-Kα radiation by a conventional $\theta$–$2\theta$ method. Rietveld refinement was performed using RIETAN-FP[28]. Crystal structure images were drawn using VESTA[29]. The actual composition of the synthesized polycrystalline samples was investigated by energy dispersive X-ray spectroscopy (EDX) on a scanning electron microscope SEM, TM-3030 (Hitachi), with Swift-ED (Oxford). The obtained compositions are listed in Table S1 (Supplemental data).

*Measurements of superconducting properties*

The superconducting properties were investigated using a superconducting quantum interference device (SQUID) magnetometer on MPMS-3 (Quantum Design). The temperature dependence of magnetic susceptibility ($4\pi\chi$) was measured after both zero-field cooling (ZFC) and field cooling (FC) with an applied field of ~10 Oe. The temperature dependences of susceptibility for all the samples are shown in Fig. S1 (Supplemental data), and the estimated $T_c$ is listed in Table S1. To estimate $J_c^{global}$ of cube-shaped samples, magnetization-magnetic field (*M-H*) loops were measured. From the obtained *M-H* loops, $J_c^{global}$ was estimated using the Bean's model[30]: $J_c = 20\Delta M / B(1-B/3A)$ (A/cm$^2$), where $A$ and $B$ are lengths determined by sample shape, and $\Delta M$ is obtained from the width of the *M-H* curve. Typical results on $J_c^{global}$ are plotted as a function of field in Fig. S2 (Supplemental data). To estimate $J_c^{local}$, remanent magnetization ($m_R$) was measured[31]. The $H_m$ dependence of $dm_R/dlogH_m$ was measured using a sequence for remanent magnetization measurements. First of all, maximum magnetic field ($H_m$) was applied to the sample, and the magnetic field was set to zero, followed by magnetization measurement, which gives $m_R$ ($H_m$). The same measurements were performed at different maximum fields and temperatures. From the $H_m$ dependence of $dm_R/dlogH_m$, $H_{p2}$ was estimated from the peak position ($H_{peak}$)[31]: $H_{peak}$ = (6-2$\sqrt{2}$)$H_{p2}$ / 7. $J_c^{local}$ was calculated from $H_{p2} = J_c^{local} r$, where $r$ is average grain size. $RE$123 powder with grain sizes close to 20 μm was prepared using micro sieves and used for remanent magnetization measurements. The grain size was estimated using ImageJ[32].

**Results and discussion**

Figure 1a shows a schematic image of the crystal structure of HEA-type $RE$123 and the concept of HEA-type site at the $RE$ site. With increasing number of $RE$ in the site, $\Delta S_{mix}$ (ideal value) increases and exceeds 1.5$R$ when five $RE$ ions are solved. Therefore, $RE$123 samples with more than four $RE$ ions are expected to be HEA-type $RE$123. In Fig. 1b, inter-grain global $J_c$ ($J_c^{global}$) estimated from the $M$-$H$ loops is plotted as a function of orthorhombicity parameter ($OP$): $OP = 2|a - b| / (a + b)$. Here, $OP$ was used to discuss the superconducting properties because, as mentioned earlier, orthorhombicity is a key structural parameter to achieve higher superconducting characteristics in the $RE$123 system[12,26]. Since $OP$ is a good scale for estimating oxygen deficiency in the $RE$123 system, the use of $OP$ in plotting superconducting properties is useful to extract the effects of $\Delta S_{mix}$ at the $RE$ site of $RE$Ba$_2$Cu$_3$O$_{7-\delta}$ on the $J_c$ characteristics in the system. According to the data in Fig. 1b, we found that a higher $J_c^{global}$s are achieved for samples containing four, six, and seven $RE$ elements. The results motivated us to study $J_c^{local}$ of $RE$123 samples with different $\Delta S_{mix}$ to clarify the effect of high configurational entropy of mixing to $J_c$ characteristics. Then, seven samples having different $\Delta S_{mix}$ and almost comparable $OP$ were selected for this study (highlighted with a square in Fig. 1b). As well, bulk nature of superconductivity of the sample was confirmed through $M$-$T$ measurements as displayed in Fig. S1 (Supplemental data). According to the number of $RE$ elements, examined samples are labeled RE-1–RE-7 (see Table I).

Through EDX analyses, we confirmed that the actual composition of the examined samples is comparable to the nominal value, as summarized in Table I. Figure 2a shows powder XRD patterns for all the samples (RE-1–RE-7). All the peaks could be indexed with the orthorhombic $RE$123 model with a space group of $Pmmm$ (#47). As displayed in Fig. 2b, no peak broadening was observed among seven samples, indicating that the increase in $\Delta S_{mix}$ does not affect crystallinity of the polycrystalline *RE*123 samples. The major peaks shifted according to the average ionic radius at the *RE* site. As mentioned earlier, however, in the *RE*123 system, *OP* is the essential parameter for superconducting properties rather than lattice constants. We, therefore, estimated lattice constants, *a* and *b*, using the Rietveld refinements. The typical refinement result with a reliability factor, $R_{wp}$ = 5.8%, for RE-5 is shown in Fig. 2c, which shows that the orthorhombic model can nicely reproduce the XRD patterns even for a HEA-type sample with five different *RE*. The estimated lattice constants and the calculated *OP* are summarized in Table I.

To perform remanent magnetization ($m_R$) measurements, powders with similar grain sizes are prepared and observed by SEM. As shown in Fig. 3, the grain size of the powders was almost homogeneous, and the estimation of the average grain size was successful using ImageJ software. The estimated average grain size for RE-1–RE-7 is summarized in Table II.

Figure 4 shows the results of the $m_R$ measurements at $T$ = 2.0, 4.2, 10.0, 20.0 K plotted in a form of $dm_R/d\log H_m$ as a function of maximum applied field ($H_m$). Basically, we observed two peaks in the plots; the first peak observed at lower fields is originating from $J_c^{global}$. Since powder samples were used for the $m_R$ measurements, $J_c^{global}$ is reasonably quite low. The second peak at higher fields is originating from $J_c^{local}$. From the peak position ($H_{peak}$), $J_c^{local}$ was calculated using the equations $H_{peak} = (6-2\sqrt{2})H_{p2} / 7$ and $H_{p2} = J_c^{local} r$, where $r$ is average grain size, as described in Materials and methods. The estimated $J_c^{local}$ at $T$ = 2.0, 4.2, 10.0, 20.0 K for RE-1–RE-7 is plotted in Fig. 5 as a function of number of *RE* elements. At 20.0 K, there is no clear difference in $J_c^{local}$ among the seven samples except for slightly higher values for RE-1 and RE-4. At lower temperatures, the difference in $J_c^{local}$ becomes remarkable. At 4.2 K, $J_c^{local}$ of RE-3, RE-4, and RE-5 is higher than that of RE-1. Furthermore, at 2.0 K, $J_c^{local}$ of RE-4 and RE-5 is clearly higher than that of RE-1, RE-2, and RE-3. Our results on $J_c^{local}$ measured for the polycrystalline *RE*123 powders suggest that optimization of $\Delta S_{mix}$ at the *RE* site can improve $J_c^{local}$ of *RE*123 materials at low temperatures.

Having compared the *OP* parameter for RE-1, RE-4, and RE-5, we found that *OP* for those samples is almost the same. Therefore, the oxygen amount would be comparable in those samples. Therefore, the improvement of $J_c^{local}$ by the increase in $\Delta S_{mix}$ observed between RE-1, RE-4, and RE-5 would be originating from local structural modification by the disordered *RE* site and chemical bonds near the *RE* site. In BiS$_2$-based *RE*(O,F)BiS$_2$ superconductors, local structure modification in the conducting BiS$_2$ layers by the increase in $\Delta S_{mix}$ was observed[33]. Therefore, we expect that structural modulation was generated in the CuO$_2$ planes by introduction of middle- or high-entropy-alloy-type *RE* site. The highly disordered *RE* site and the bonds would modify electronic states and microscopic characteristics of superconductivity in the samples. The lower $J_c^{local}$ observed for RE-6 and RE-7 than that for RE-4 or RE-5 may be caused by huge disordered states. Although we have no clear scenario of the enhancement of pinning properties by the increase in $\Delta S_{mix}$, our current results encourage further studies on the relationship between configurational entropy of mixing and critical current properties in the *RE*123 materials. Synthesis of single crystals or thin films of HEA-type REBa$_2$Cu$_3$O$_{7-\delta}$ and investigation on local structure modulations and microscopic characteristics of superconductivity are needed to clarify the HEA effects in *RE*123.

**Conclusion**

Here, we reported the improvement of $J_c^{local}$ in *RE*Ba$_2$Cu$_3$O$_{7-\delta}$ by the increase in $\Delta S_{mix}$ at the *RE* site. Polycrystalline *RE*123 samples with different $\Delta S_{mix}$ were synthesized by solid-state reaction. Through characterization of structural (lattice constants and *OP*), compositional, and superconducting properties ($T_c$ and shielding fraction), seven samples labeled RE-1–RE-7 were chosen, and remanent magnetization measurements were performed on those samples. At higher temperatures ($T$ = 20.0 K), clear difference in $J_c^{local}$ was not observed. At lower temperatures ($T$ = 2.0 and 4.2 K), higher $J_c^{local}$ was observed for RE-4 and RE-5 with a higher $\Delta S_{mix}$ as compared to that for RE-1, RE-2, and RE-3 with zero or low entropy of mixing at the *RE* site. Although the results of the current work showed the merit of high-entropy alloying at lower temperatures only, there should be optimal conditions on constituent *RE* element, mixing ratio, and $\Delta S_{mix}$, which will achieve higher $J_c^{local}$ at higher temperatures as well. If the trial was achieved, the HEA concept can be applied to all *RE*123 practical materials to additionally improve their critical current properties.

**Table I. Compositional, structural, and superconducting properties of $RE$Ba$_2$Cu$_3$O$_{7-\delta}$ samples examined in this study.**

| Label | $RE$ site (EDX) | $\Delta S_{mix}$ ($RE$) | $a$ (Å) | $b$ (Å) | $OP$ | $T_c$ (K) |
|---|---|---|---|---|---|---|
| RE-1 | Y | 0 | 3.81375(10) | 3.8804(2) | 0.0173 | 92.9 |
| RE-2 | Y$_{0.57}$Nd$_{0.43}$ | 0.68$R$ | 3.85007(13) | 3.9126(2) | 0.0161 | 92.2 |
| RE-3 | Y$_{0.39}$Sm$_{0.30}$Eu$_{0.31}$ | 1.09$R$ | 3.84149(10) | 3.9038(2) | 0.0161 | 93.2 |
| RE-4 | Y$_{0.32}$Sm$_{0.22}$Eu$_{0.26}$Dy$_{0.20}$ | 1.37$R$ | 3.8340(2) | 3.89927(10) | 0.0169 | 92.8 |
| RE-5 | Y$_{0.24}$Sm$_{0.15}$Eu$_{0.17}$Dy$_{0.16}$Ho$_{0.28}$ | 1.58$R$ | 3.8290(2) | 3.89593(8) | 0.0173 | 93.0 |
| RE-6 | Y$_{0.16}$Sm$_{0.14}$Eu$_{0.15}$Dy$_{0.14}$Ho$_{0.14}$Yb$_{0.27}$ | 1.76$R$ | 3.8163(3) | 3.8858(12) | 0.0180 | 92.0 |
| RE-7 | Y$_{0.15}$Sm$_{0.10}$Eu$_{0.12}$Dy$_{0.12}$Ho$_{0.17}$Yb$_{0.22}$Lu$_{0.12}$ | 1.91$R$ | 3.81933(14) | 3.89045(10) | 0.0185 | 91.4 |

**Table II. Average grain area and size of the $RE$Ba$_2$Cu$_3$O$_{7-\delta}$ powders used for the estimation of $J_c^{local}$.**

| Label | Average grain area (μm$^2$) | Average grain size (μm) |
|---|---|---|
| RE-1 | 472.23 | 21.7 |
| RE-2 | 508.567 | 22.6 |
| RE-3 | 420.793 | 20.5 |
| RE-4 | 321.049 | 17.9 |
| RE-5 | 427.58 | 20.7 |
| RE-6 | 438.918 | 21.0 |
| RE-7 | 437.66 | 20.9 |

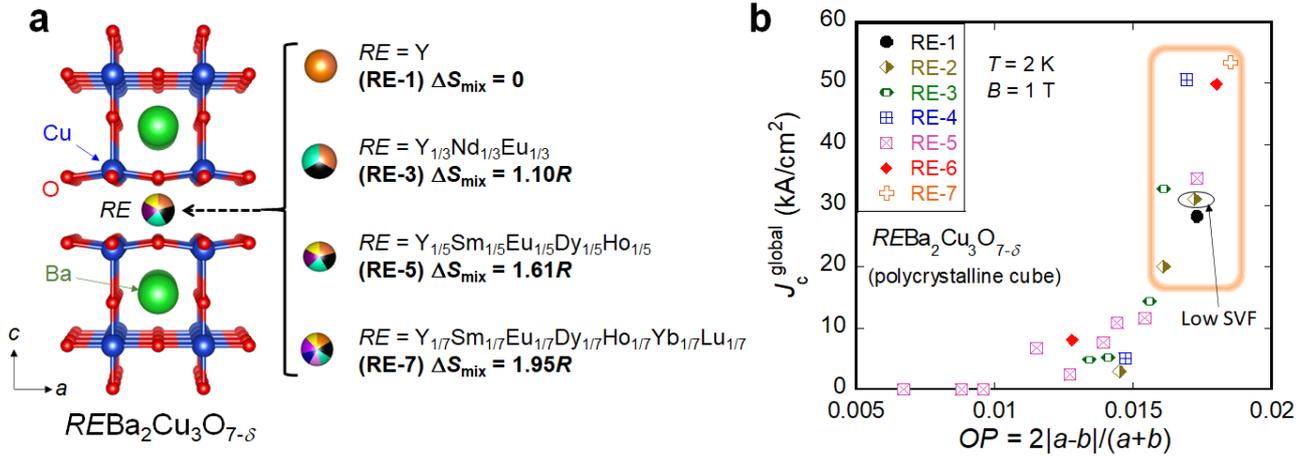

**Fig. 1. Structural and in-field $J_c^{global}$ characteristics of polycrystalline $RE$Ba$_2$Cu$_3$O$_{7-\delta}$. a.** Schematic images of crystal structure of $RE$Ba$_2$Cu$_3$O$_{7-\delta}$ and high-entropy alloying at the $RE$ site. The $\Delta S_{mix}$ was calculated according to the nominal values of $RE$ concentration. **b.** Magnetic $J_c^{global}$ ($T$ = 2 K, $B$ = 1 T) as a function of orthorhombicity parameter ($OP = 2\,|a - b|\,/\,(a + b)$). In this study, we examined $J_c^{local}$ for samples surrounded by an orange square in the figure 1b. Because of low shielding volume fraction (SVF) estimated from temperature dependence of susceptibility, one RE-2 sample containing Y and Sm was excluded in this study. Some of the $J_c^{global}$ data have already been published in Ref. 26.

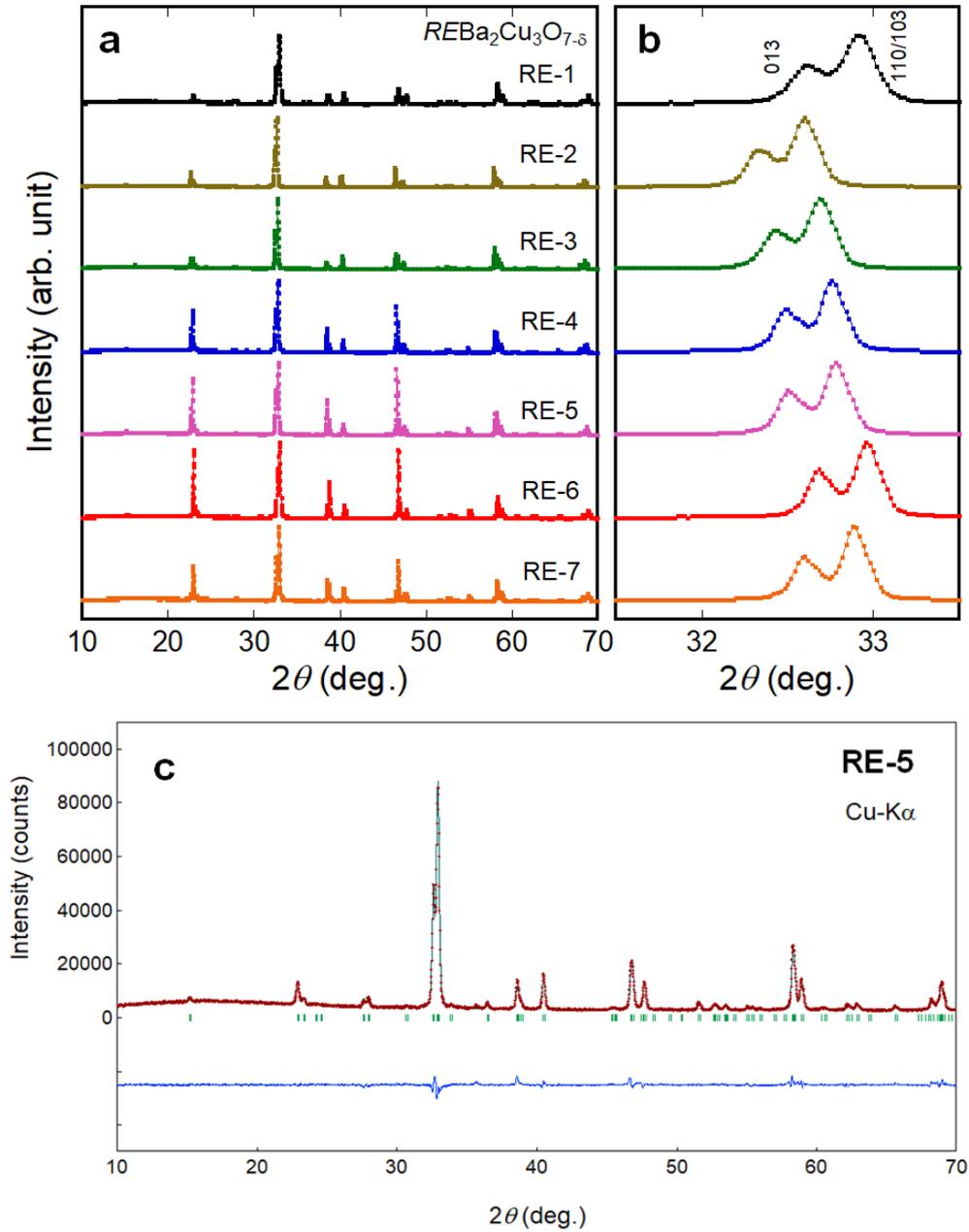

Fig. 2. XRD patterns of *RE*Ba$_2$Cu$_3$O$_{7-\delta}$ samples. a. Powder XRD patterns of all the examined samples, RE-1–RE-7. b. Zoomed plots near the (013) peak for RE-1–RE-7. c. Rietveld refinement result for RE-5.

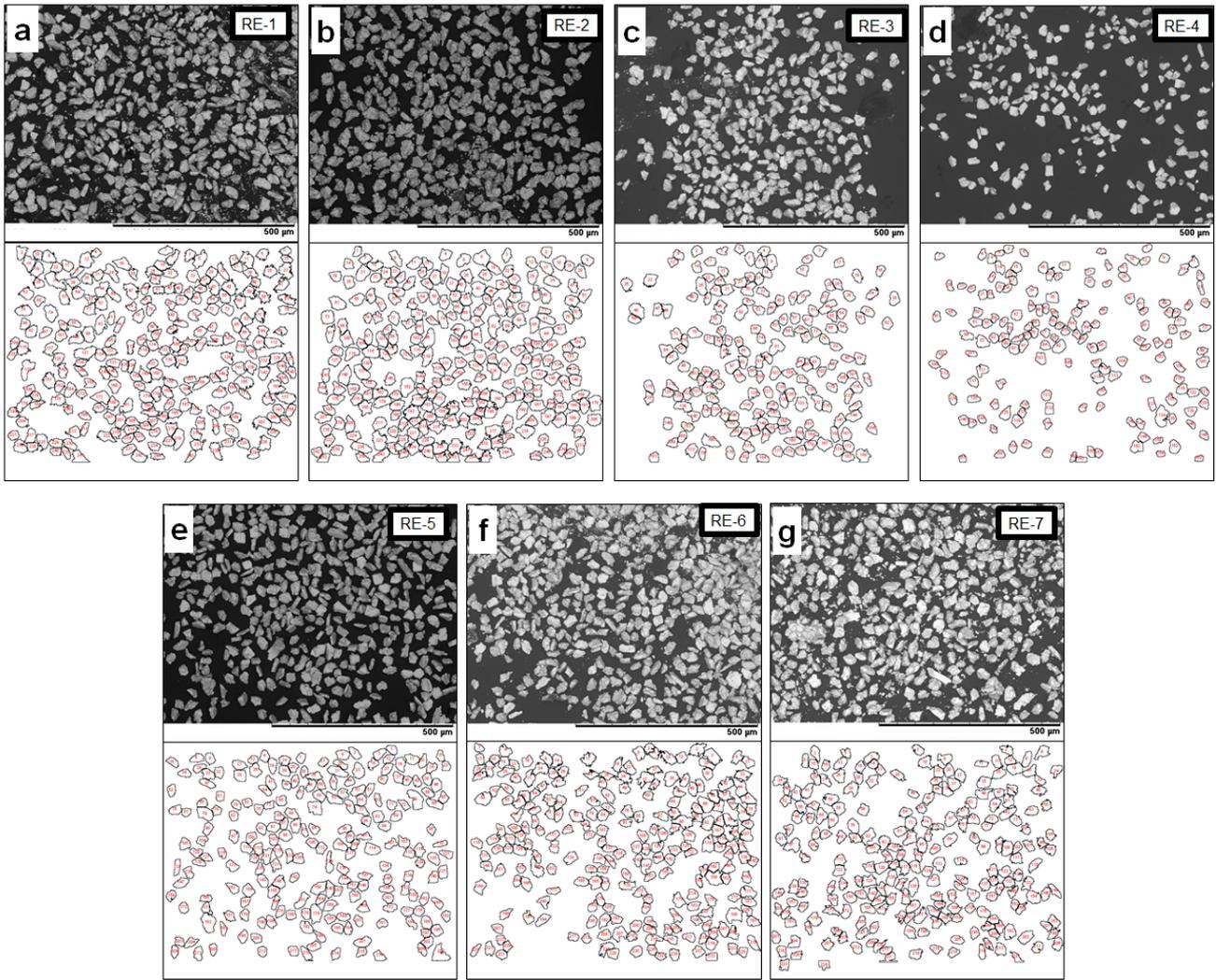

**Fig. 3. Grain characterization. a–g.** SEM images of $RE$Ba$_2$Cu$_3$O$_{7-\delta}$ grains (upper panels) and grain-size analyses (lower panels) for all the samples, RE-1–RE-7. SEM images were analyzed using ImageJ software[32].

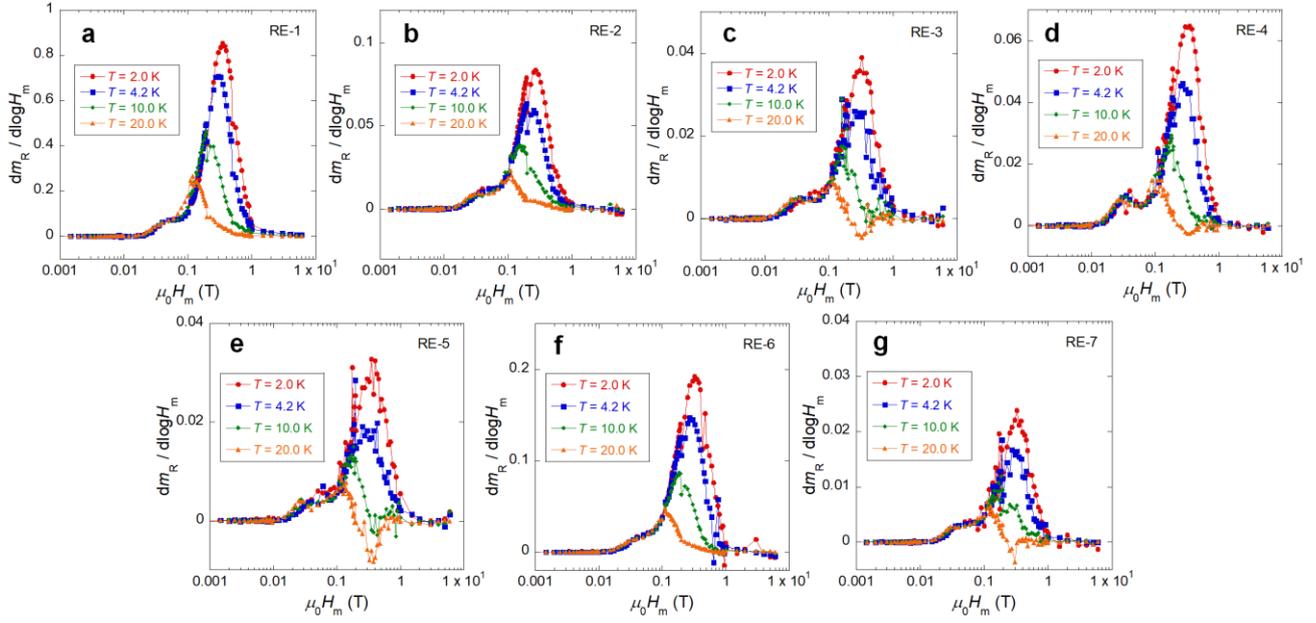

**Fig. 4. Results of remanent magnetization measurements. a–g.** Maximum magnetic field ($\mu_0 H_m$) dependences of the $H_m$ derivative of remanent magnetization ($m_R$), in a form of $dm_R/d\log H_m$, for RE-1–RE-7.

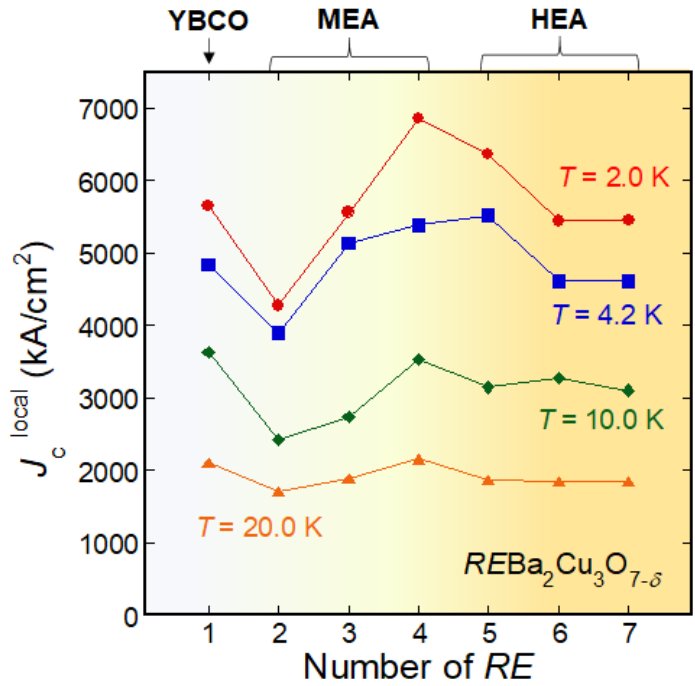

**Fig. 5. Estimated $J_c^{local}$ as a function of number of $RE$ elements.**

**Acknowledgements**

The authors thank Y. Goto, R. Kurita and O. Miura for supports in experiments and discussion. This work was partly supported by JSPS-KAKENHI (18KK0076) and Tokyo Metropolitan Government Advanced Research (H31-1).

**Authors' contributions**

Y.M. designed the research. Y.S. and A.Y. prepared the samples and characterized. Y.S. carried out *M-T* and *M-H* measurements. A.Y. carried out remanent magnetization measurements. A.Y. and Y.M. analyzed remanent magnetization data. A.Y. and Y.M. wrote the manuscript. All the authors made contributions to writing the manuscript.

**Conflict of interest**

The authors declare that they have no conflict of interest.

# Supplemental data

Table S1. Information of examined samples: nominal composition, orthorhombicity parameter (*OP*), transition temperature ($T_c$), and $J_c^{global}$ ($T$ = 2 K, $B$ = 1 T) estimated from magnetization-magnetic field (*M-H*) curve and the Bean's model. Some of the data in the table have already been published in Y. Shukunami et al., Phys. C 572, 1353623 (2020).

| *RE* site composition (nominal) | $OP = 2|a-b|/(a+b)$ | $T_c$ (K) | $J_c^{global}$ (kA/cm$^2$) (2 K, 1 T) |
|---|---|---|---|
| Y | 0.0173 | 92.9 | 28.3 |
| Y$_{1/2}$La$_{1/2}$ | 0.0145 | 88 | 3.0 |
| Y$_{1/2}$Nd$_{1/2}$ | 0.0161 | 92.2 | 20.1 |
| Y$_{1/2}$Sm$_{1/2}$ | 0.0172 | 92.7 | 31.2 |
| Y$_{1/3}$La$_{1/3}$Nd$_{1/3}$ | 0.0141 | 90.5 | 5.2 |
| Y$_{1/3}$La$_{1/3}$Sm$_{1/3}$ | 0.0134 | 91.5 | 4.8 |
| Y$_{1/3}$Nd$_{1/3}$Sm$_{1/3}$ | 0.0156 | 92.4 | 14.4 |
| Y$_{1/3}$Sm$_{1/3}$Eu$_{1/3}$ | 0.0161 | 93.2 | 32.9 |
| Y$_{1/4}$La$_{1/4}$Nd$_{1/4}$Sm$_{1/4}$ | 0.0147 | 92.3 | 5.1 |
| Y$_{1/4}$Sm$_{1/4}$Eu$_{1/4}$Dy$_{1/4}$ | 0.0169 | 92.8 | 50.6 |
| Y$_{1/5}$La$_{1/5}$Nd$_{1/5}$Sm$_{1/5}$Eu$_{1/5}$ | 0.0115 | 92.9 | 6.8 |
| Y$_{1/5}$La$_{1/5}$Nd$_{1/5}$Sm$_{1/5}$Gd$_{1/5}$ | 0.0127 | 92.2 | 2.4 |
| Y$_{1/5}$La$_{1/5}$Nd$_{1/5}$Eu$_{1/5}$Gd$_{1/5}$ | 0.0139 | 93 | 7.7 |
| Y$_{1/5}$La$_{1/5}$Sm$_{1/5}$Eu$_{1/5}$Gd$_{1/5}$ | 0.0144 | 93.4 | 10.9 |
| Y$_{1/5}$Nd$_{1/5}$Sm$_{1/5}$Eu$_{1/5}$Gd$_{1/5}$ | 0.0154 | 93.4 | 11.7 |
| Y$_{1/5}$Sm$_{1/5}$Eu$_{1/5}$Dy$_{1/5}$Ho$_{1/5}$ | 0.0173 | 93 | 34.6 |
| Y$_{1/5}$La$_{1/5}$Nd$_{1/5}$Sm$_{1/5}$Eu$_{1/5}$ | 0.0040 | - | - |
| Y$_{1/5}$La$_{1/5}$Nd$_{1/5}$Sm$_{1/5}$Gd$_{1/5}$ | 0.0067 | - | - |
| Y$_{1/5}$La$_{1/5}$Nd$_{1/5}$Eu$_{1/5}$Gd$_{1/5}$ | 0.0088 | - | - |
| Y$_{1/5}$La$_{1/5}$Sm$_{1/5}$Eu$_{1/5}$Gd$_{1/5}$ | 0.0096 | - | - |
| Y$_{1/6}$La$_{1/6}$Nd$_{1/6}$Sm$_{1/6}$Eu$_{1/6}$Gd$_{1/6}$ | 0.0128 | 93.1 | 8.1 |
| Y$_{1/6}$Sm$_{1/6}$Eu$_{1/6}$Dy$_{1/6}$Ho$_{1/6}$Yb$_{1/6}$ | 0.0180 | 92 | 49.9 |
| Y$_{1/7}$Sm$_{1/7}$Eu$_{1/7}$Dy$_{1/7}$Ho$_{1/7}$Yb$_{1/7}$Lu$_{1/7}$ | 0.0185 | 91.4 | 53.5 |

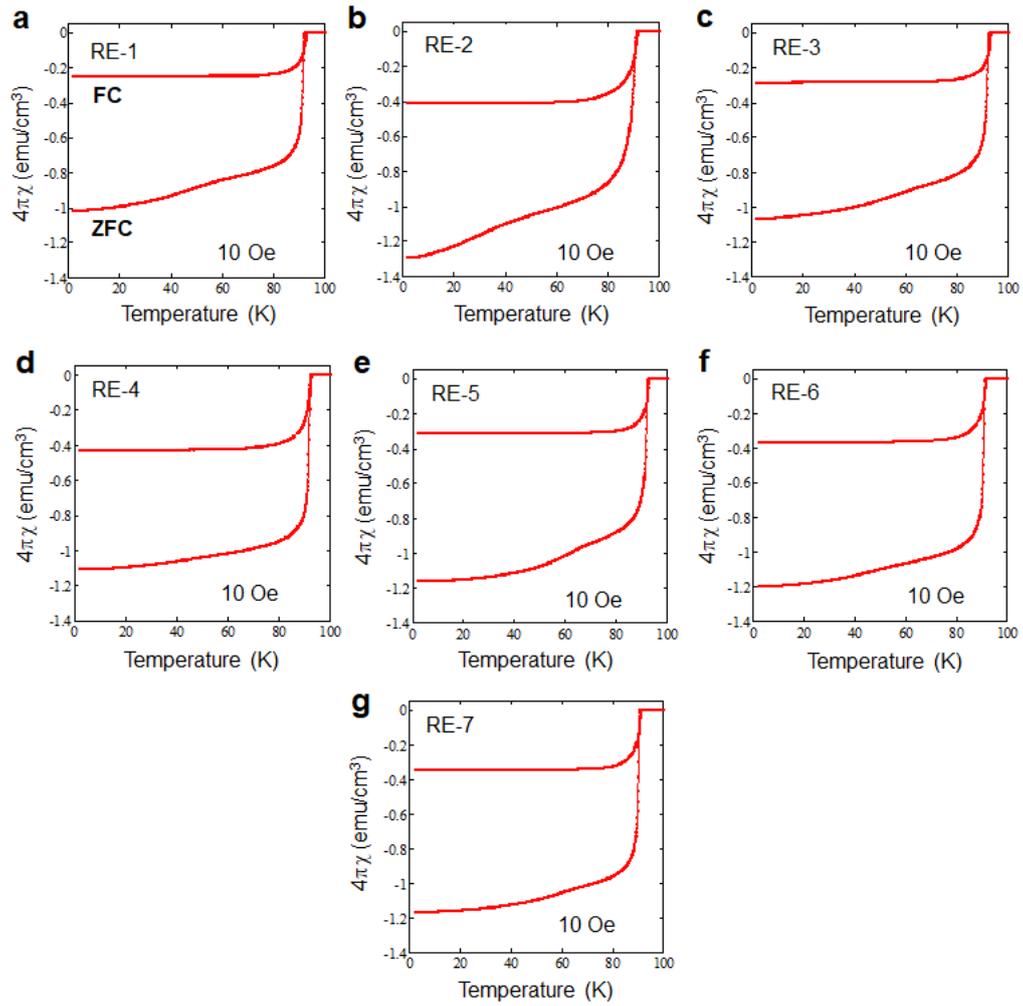

Fig. S1. Temperature dependence of magnetic susceptibility ($4\pi\chi$) after zero-field cooling (ZFC) and field cooling (FC) with an applied field of ~10 Oe for sample RE-1–RE-7.

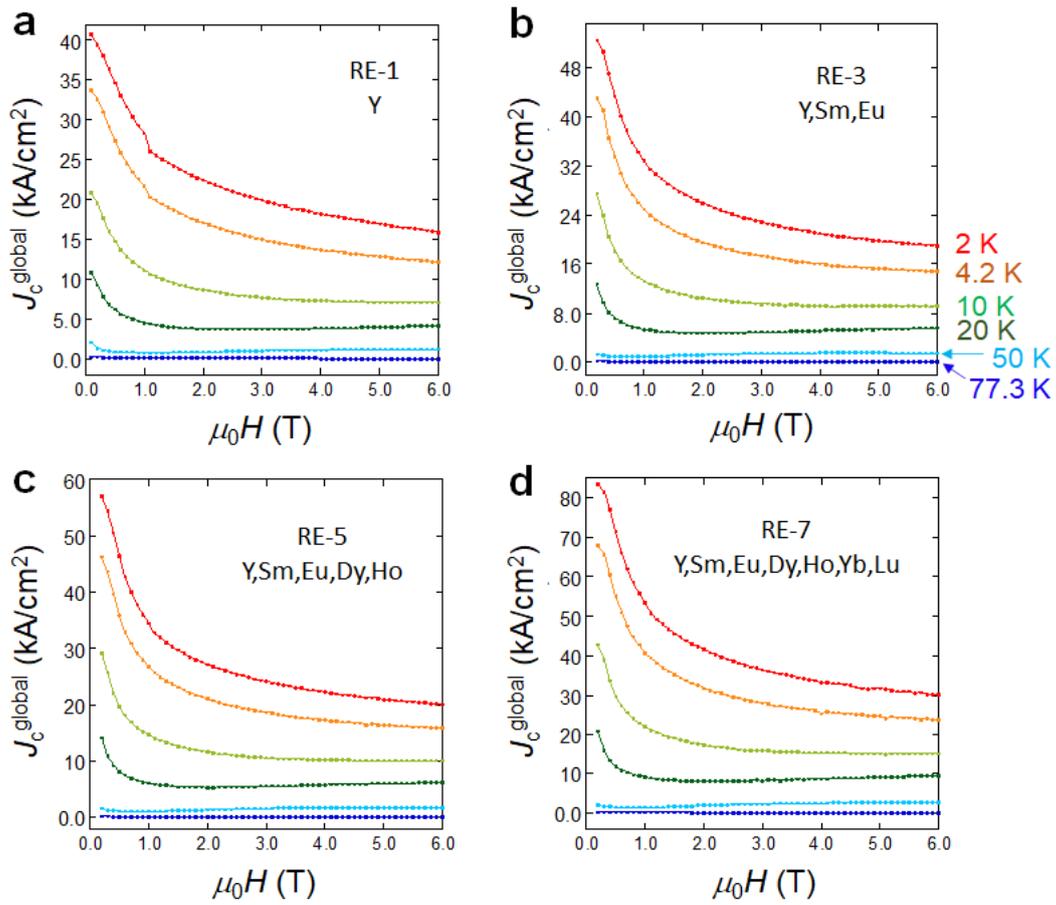

Fig. S2. Typical data of $J_c^{global}$ as a function of magnetic field for sample RE-1, RE-3, RE-5, and RE-7.